\documentclass[aps,superscriptaddress]{revtex4-2}
\usepackage{amsmath,amssymb}
\usepackage{graphics,graphicx,subfigure}
\usepackage{dcolumn,bm}
\usepackage{psfrag}
\usepackage{color}
\usepackage{multirow}
\usepackage{braket}
\newcommand{\cu}
{\affiliation{Vidyasagar College, 39 Sankar Ghosh Lane, Kolkata 700006}}
\newcommand{\be}
{\begin{equation}}
\newcommand{\ee}
{\end{equation}}

\begin{document}

\title{A Poor Agent and Subsidy: An investigation through CCM Model}
\author{Sanchari Goswami}
\cu
\maketitle
\begin{center}
	Abstract
\end{center}
	In this work, the dynamics of agents below a \textit{threshold line} 
	in some modified CCM type kinetic wealth exchange models are studied. 
	These agents are eligible for subsidy as can be seen in any real economy. 
	An interaction is prohibited if both of the interacting agents' wealth fall below the 
	threshold line. A walk for such agents can be conceived in the abstract Gain-Loss Space(GLS) 
	and is macroscopically compared to a lazy walk. The effect of giving subsidy once to such agents 
	is checked over giving repeated subsidy from the point of view of the walk in GLS. It is seen that the 
	walk has more positive drift if the subsidy is given once. The correlations 
	and other interesting quantities are studied.  

\section{Introduction}

In any economy the distribution of wealth $P(m)$ among individuals follows a pattern for large values 
of wealth $m$, to be specific, it decays 
as $P(m) \sim  m^{-(1+\nu)}$ for large $m$ where $\nu$ is called the Pareto exponent \cite{Pareto}. Pareto exponent usually varies between $1$
and $3$ \cite{Mandelbrot:1960, EIWD, EWD05, ESTP, SCCC, Yakovenko:RMP, datapap,DraguYakov0}. 
A number of models have been proposed
to reproduce the observed features of an economy \cite{marjitIspolatov, Dragulescu:2000,
Chakraborti:2000,Chatterjee:rev,Chatterjee:2010}. One important objective of several such econophysical models 
is to reproduce the Pareto tail 
in the wealth/income distribution. Some of the models were inspired by the kinetic theory of 
gases which derives the average macroscopic behaviour from the microscopic interactions among molecules. 
In these models traders/agents are treated as molecules of gas. A typical trading process between two such 
traders/agents maintaining local conservation of wealth 
can be compared to an interaction between two gas molecules maintaining local energy conservation in gas. 
These models follow a microcanonical description, i.e., the total wealth 
is a conserved quantity. 
Several such models are studied \cite{EIWD, EWD05, Toscani2018} where debt is allowed for a trader/agent. However, in our case  
we consider no agent can end up with a negative wealth, i.e., debt is not allowed in a trading. 

Thus, if there are two agents $i$ and $j$ who before taking part in the trading had wealth 
$m_i(t)$ and $m_j(t)$ respectively at time $t$, will have wealths according to the following relations 
at the next time step $t+1$:\\
\begin{align}\label{mimj}
\begin{split}
m_i(t+1) = m_i(t) + \Delta m; \\
m_j(t+1) = m_j(t) - \Delta m.\\
\end{split}
\end{align}

There are several other models of the wealth distribution which do not consider the kinetic theory concept.  
In \cite{Bouchaud-Mezard}, a very simple model of economy was discussed, where the time evolution
was described by an equation involving exchange between individuals and
random speculative trading in such a way that under an arbitrary change of monetary units 
the fundamental symmetry of the economy
is obeyed.
A mean-field limit of this
equation was investigated there and the distribution of wealth came out to be of the Pareto type.
Another model is the Lotka-Volterra 
model which is again a kind of mean field model where wealth of an agent at a particular time 
depends on her/his wealth in the previous step as well as the average 
wealth of all agents \cite{Solomon,Malcai}. Apart from these, there are other 
models which depend on stochastic processes \cite{Garl,Sornette}. The main problem in the last two type 
models is that here wealth 
exchange between agents is not allowed 
and therefore cannot be realized as a real trading process. Although in \cite{Bouchaud-Mezard}, wealth exchange is 
considered, according to the authors, it is again not a fully realistic one, as mean field concept is used.
In some models, 
instead of considering binary collision-like trading, just as in case of a rarefied classical gas, simultaneous
multiple interactions are taken into account to model a socio-economic phenomena 
in a multi-agent system \cite{Toscani}. \\

In the gas-like models, the wealth exchange between agents follow the same rule as energy exchange between two gas 
molecules in kinetic theory; that is why they are called \textit{kinetic wealth exchange models}. 
Bachelier in his $1900$ PhD. thesis developed a 'theory  of speculation' \cite{Bachelier1900}, 
where he suggested a practical connection between stochastic theory 
and financial analysis.
The idea that velocity distribution for gas molecules and income distribution for agents 
can be compared was first addressed in \cite{SahaSrivastava}, 
although no specific reason behind this was addressed. The 
first simplest conservative model of this kind was proposed by  
Dragulescu and Yakovenko (DY model) \cite{Dragulescu:2000}. In that model, $N$ agents 
randomly exchange wealth pairwise keeping the total wealth $M$ constant. 
It is shown that the steady-state ($t \rightarrow \infty$) wealth there follows a Boltzmann-Gibbs distribution:
$P(m)=(1/T)\exp(-m/T)$; $T=M/N$ ~\cite{Chatterjee:rev}.\\

A modification to this model considering the fact that agents save a definite fraction of their wealth $\lambda$ before 
taking part in any trading, termed as \textit{saving propensity}, was addressed first by Chakraborti 
and Chakrabarti~\cite{Chakraborti:2000} 
(CC model).    
This results in a wealth 
distribution close to Gamma distributions~\cite{Patriarca:2004,Repetowicz:2005} and is seen
to fit well to empirical data for low and middle wealth regime of an economy \cite{datapap}.
Later, a  model was proposed by Chatterjee et. al. \cite{Chatterjee:2004} 
(CCM model) where distributed saving propensities were assumed for individuals. 
The importance of the model is that it led to a wealth distribution 
with a Pareto-tail. Apart from wealth distribution, people often study network like features in these models
\cite{tummi2, Gabaix, manna1,goswami1}, a few of which address preferential interaction between agents. In \cite{goswami1}
it was considered that two agents will interact with more probability if their wealths are ``close'' or if
they have interacted before.\\

In a real economy, however, this preferential interaction often
depend on some other factors. Restriction in interaction may arise in some situations as we have recently 
seen in the Pandemic situation. This type of restricted interaction is studied in \cite{Toscani2}. 
Also during the economic crisis in Argentina during $2000-2009$ another restricted interaction was studied in \cite{Ferrero}. 
There may be restriction in interaction for other reasons too. 
It is known that 
a poverty line exists in any economy \cite{gov_report, brady}. In various cases, the poverty line is estimated 
near $40\%-60\%$ of the median of income\cite{brady}. People below the poverty line often get
subsidy from the Government. Also, it is a general notion that a person feels insecure if her/his 
wealth falls below a specific level, may be termed as a \textit{threshold line}. 
In this work, a \textit{threshold line} is introduced in an otherwise CCM like model, which is below 
the defined poverty line in 
an economy. The wealths of $N$ agents 
are chosen from the uniform distribution and the total wealth is taken as $M$. 
The agents whose wealth are assigned below the threshold line are Below Threshold Line (BTL) agents. The subsidy 
is given to the BTL agents in such a way that it can just promote 
the agents above threshold line. 
Once an agent is marked a BTL one, he/she remains eligible for subsidy always.  
However, those who are 
above the threshold line at the beginning are not getting any subsidy 
even if their wealth fall below  the threshold 
line after a certain number of interactions. 
Also an interaction will not occur at all if both the interacting agents are having wealth below the threshold line
because of human psychology of insecurity. \\

In some earlier works to study the dynamics of the transactions, a walk was conceived for the
agents in an abstract $1$-D Gain-Loss space (GLS) \cite{chattsen, goswami2}. 
The corresponding walk was compared to a biased random walk. 
In this work, we compare the walk of a tagged agent $k$ to a lazy random walk for different values of 
saving propensity $\lambda_k$. The difference from earlier work is that, here a 
tagged agent is a BTL one and except from moving Right/Left, the walker 
may stay put to its position in the GLS if the interaction does not occur at all. It is seen that average distance 
travelled in the GLS, i.e.,
$\langle x \rangle = 0$ for some $\lambda_k=\lambda_{k}^*$. The value of $\lambda_{k}^*$ is slightly different from what 
we found in \cite{goswami2}. The wealth distribution and several other features 
of the lazy walker are studied in this context. The objective of this study is to check whether there is any difference 
in the agent's upliftment in the wealth space
if the subsidy is given repeatedly to a BTL agent or only once.\\

\section{Model Description}
We consider CCM model with $N=256$ agents. The total money $M$ is distributed randomly among the agents. 
The key feature of CCM model is that here the saving propensities of agents are chosen from a uniform distribution. 
The wealth exchange between two traders $i$ and $j$ can be represented as:\\
\begin{align}\label{mimj_CCM}
\begin{split}
m_i(t+1)=\lambda_i m_i(t) + \epsilon_{ij} \left[(1-\lambda_i)m_i(t) + (1-\lambda_j)m_j(t)\right],\\
m_j(t+1)=\lambda_j m_j(t) + (1-\epsilon_{ij}) \left[(1-\lambda_i)m_i(t) + (1-\lambda_j)m_j(t)\right];
\end{split}
\end{align}
Here $\lambda_i, \lambda_j$ are the saving propensities of agents $i,j$ respectively and $\epsilon_{ij}$ is a random 
fraction related to stochastic nature of a trading process.
In addition to this regular interaction, a threshold line $m_{L}$ is proposed here. 
We assumed the poverty line near $40\%$ of the average wealth of the economy as indicated in \cite{brady} 
and the threshold line is chosen below that. 
At the time of wealth assignment if 
an agent is found to be below the line she/he will be marked as BTL and a 
subsidy is assigned. 
As the wealths of agents are assigned from a uniform distribution the median is same as the average wealth.
This means a certain fraction of the people will get subsidy. 
However, during course of interaction, an agent who is not a BTL one, falls below the threshold 
line she/he is not eligible for subsidy. Two types of models are studied here as follows:

\begin{itemize}
 \item Model A : For this, the BTL agents are stamped as ``BTL'' at the time of wealth assignment and 
 the subsidy equal to $m_L$ is given to the BTL agents at the beginning of each configuration. 
 \item Model B : In this, again, some are stamped ``BTL'' at the beginning but 
the subsidy equal to $m_L$ is given to them at each Monte Carlo (MC) step,
where one MC step consists of $\frac{N(N-1)}{2}$ interactions. This means if a BTL agent goes above the threshold line 
after one or few MC steps, she/he is still eligible for subsidy (exactly as in any caste based system).
\end{itemize}
 
In both the cases, the subsidy given promotes the BTL agent above the threshold line $m_L$, if the agent is below that. 
At every interaction, the wealths of the interacting agents are checked. An 
interaction is prohibited only if 
both of the interacting agents fall below the threshold line. In subsequent interactions, however, 
there is always a chance that such an agent is promoted above the line.
The stationary state is obtained after a typical relaxation time and the distribution of wealth and the walk 
in the GLS are studied. It is to be noted that the subsidy here is given from the tax payed by the 
people above the threshold line. In this way, the economy is 
closed and total wealth remains conserved. 

\begin{figure}[!h]
\centering
\includegraphics[trim=5cm 0cm 0cm 0cm, width=6cm, angle=-90,clip]{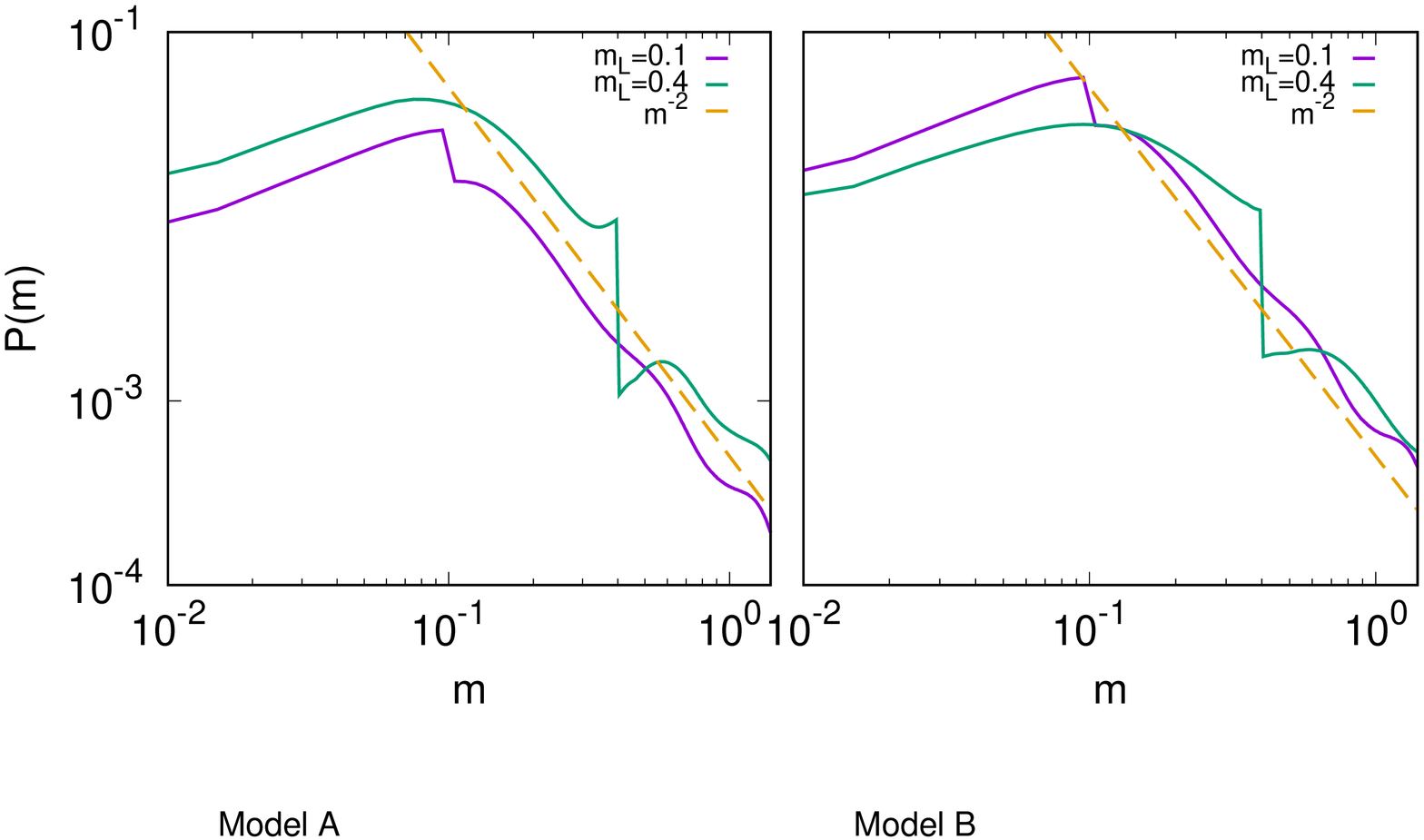}
\caption{Overall wealth distribution $P(m)$ for model A (Left) and model B (right) for $m_L=0.1$ (violet), $0.4$ (green). The Pareto exponent 
$\nu$ is found to be close to $1$.}
\label{mnydist_overall}
\end{figure}

\begin{figure}[!h]
\centering
\includegraphics[trim=5cm 0cm 0cm 0cm, width=6cm, angle=-90]{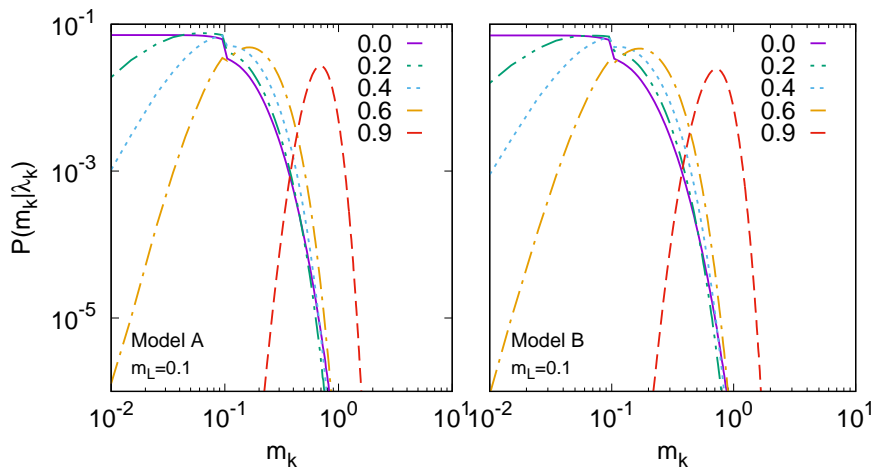}
\caption{Left: Wealth distribution for model A for $m_L=0.1$ for $\lambda_k=0.0$(violet),$0.2$(green),$0.4$(blue),$0.6$(yellow),
$0.9$(red), 
Right: Wealth distribution for model B for same $m_L=0.1$ and same $\lambda_k$ s. Both 
are for a tagged BTL agent. The plots indicate higher probability close to $m_L$.}
\label{mnydist}
\end{figure}

\section{Agent Dynamics}
Although the actual form of wealth distribution 
$P(m)$ depends on the form of saving propensity distribution, there is one thing common for all. 
The Pareto tail is present whatever be the form of the saving propensity distribution; the only difference is in the value of 
the exponent. Here, the saving propensities of all the agents are 
chosen from a uniform distribution which is the simplest one. 
Although in this work, we are actually interested in the dynamics of a tagged agent, 
the behaviour of overall distribution of wealth $P(m)$ is also of great importance. 
The overall wealth distribution shows the Pareto exponent to be roughly $1$ as in case of conventional CCM model 
\cite{Chatterjee:rev} except from the fact that near $m_L$ there is a sharp change in the profile. This behaviour 
can be understood by studying individual agent's wealth distribution which will be addressed in the next section.   
The wealth distribution $P(m)$ is shown in Fig. \ref{mnydist_overall} 
for model A (Left) and model B (right) for two different $m_L$. 


We perform numerical simulation for a system of $N$ agents
and look for the dynamics of a tagged agent who is a BTL one with a predefined saving propensity.
As we have stated earlier we use two models, namely A and B, for assigning subsidy. 

\begin{figure}[!h]
\centering
\includegraphics[trim=5cm 0cm 0cm 0cm, width=6 cm, angle=-90,clip]{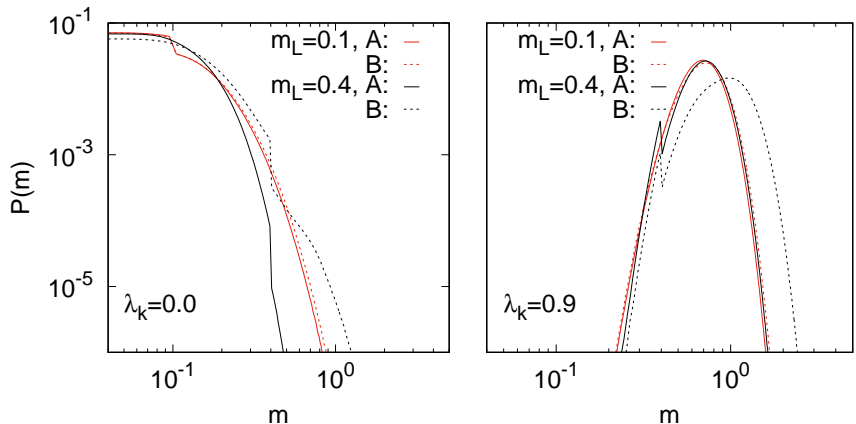}
\caption{Left: Wealth Distribution of the BTL tagged agent for both models A and B for $\lambda_k=0.0$. 
Right: Same for $\lambda_k=0.9$. The data are shown 
for $m_L=0.1$(red), $0.4$ (black). The plots indicate that for model B the wealth distribution 
extends upto a larger $m$ as we are repeating the wealth assignment. The effect is more prominent for larger values 
of $m_L$.}
\label{mnydist_compare}
\end{figure}

\subsection{Wealth Distribution}
The distribution of wealth $P(m_k|\lambda_k)$ for different $\lambda_k$ of the tagged BTL agent are shown in 
Fig. \ref{mnydist} for both the models A and B for $m_{L}=0.1$. For both the cases the nature of wealth 
distribution is similar to what observed in \cite{chattsen} and earlier \cite{Chatterjee:rev, Chatterjee:2004} 
with the exception that 
as the tagged agent is a BTL one and she/he is assigned a wealth equal to $m_L$, 
there is a higher probability near $m_L$ than what observed earlier. For small values of $\lambda_k$ the agent has a 
small amount of wealth compared to the average wealth of the economy and for higher $\lambda_k$ the wealth possessed by 
the agent is comparable to the average wealth. It is to be noted that as our agent is a BTL one, the
average wealth she/he possessed is smaller compared to that predicted by the usual CCM model.  
As we increase $m_L$ to higher values, higher $m$ is more probable. 
The models A and B show similar distribution for all $\lambda_k$ if $m_L$ 
is low but their nature is different for higher $m_L$. It is seen that 
for higher threshold line the distribution for model B shifts to higher $m$ compared to model A. This can 
be understood easily. For model B, we are assigning the subsidy to BTL agents at every MC steps without checking 
whether they are below the threshold line or not. Therefore, higher value of wealth is more probable.  
This can also be realized from another aspect of the CCM model. When we set a higher $m_L$, that means 
a larger number of agents is below that line compared to the case of a smaller $m_L$. That means we are 
moving closer to the usual CCM picture where we choose an agent irrespective of the initial wealth possessed by her/him. 
These are shown in Fig. 
\ref{mnydist_compare} for $m_L=0.1$ and $m_L=0.4$ for $\lambda_k=0.0, 0.9$.

\begin{figure}[!h]
\includegraphics[width=9 cm, angle=-90]{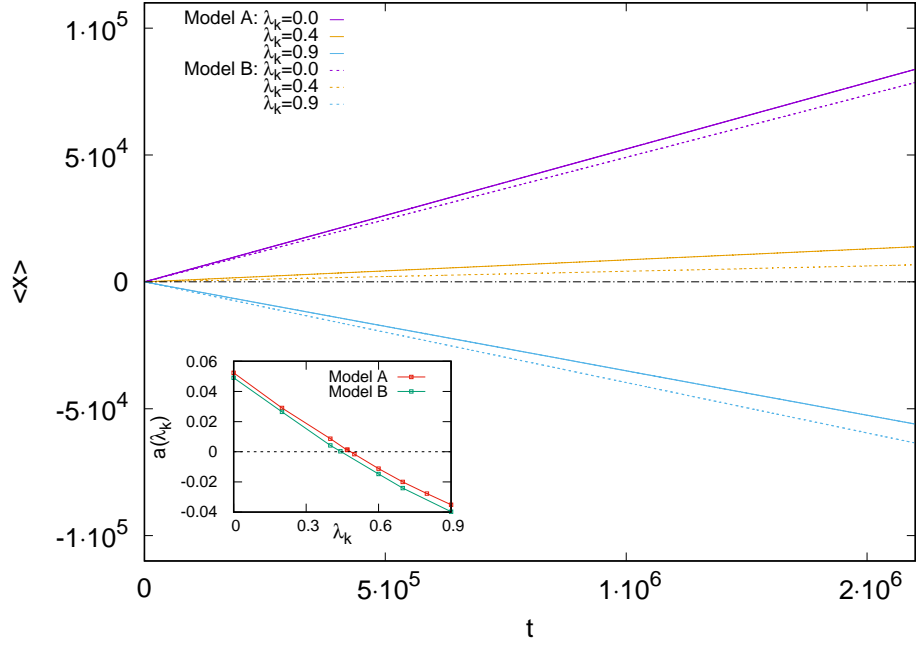}
\caption{Variation of $\langle x \rangle$ against $t$ for $\lambda_k = 0.0$ (violet),$0.4$ (yellow) and $0.9$ (blue) for the 
BTL tagged agent walker for both models A (solid lines) and B (dotted lines) for $N = 256$ for $m_L=0.1$. 
Inset shows the plot of slope $a(\lambda_k)$ as a function of $\lambda_k$ obtained from the slopes 
of $\langle x \rangle$ versus $t$ plot for the walker for both the models A and B.}
\label{slopeAB}
\end{figure}

\begin{figure}[!h]
\centering
\includegraphics[trim=5cm 0cm 0cm 0cm,width=6cm, angle=-90,clip]{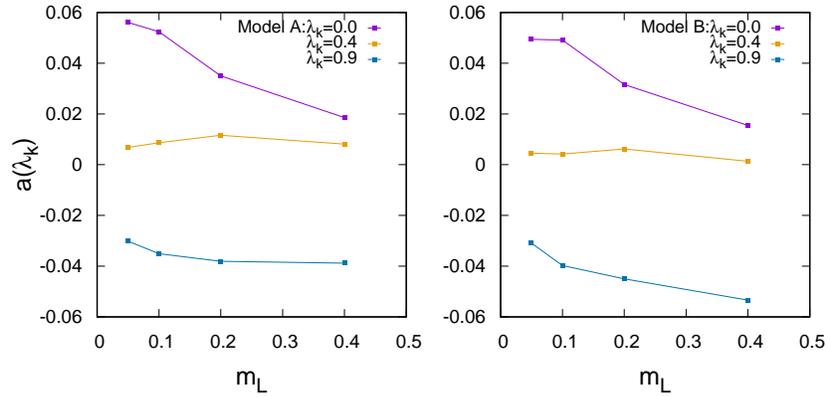}
\caption{Variation of $a(\lambda_k)$ against $m_L$ for $\lambda_k = 0.0$ (violet),$0.4$ (yellow) and $0.9$ (blue) for the 
BTL tagged agent walker for both models A (Left) and B (Right) for $N = 256$.}
\label{slopevariation_mL}
\end{figure}

\section{Walk in the GLS: Comparison to a Lazy Walker}
To investigate the dynamics of this model in more detail at the microscopic level,
one may conceive a walk for the agents in the GLS.
It is well known that the usual CCM walk can be compared to a biased 
random walk (BRW) whose forward bias decreases as we increase $\lambda$ 
from zero. The walk has no bias at a particular $\lambda_k=0.469$ and then it decreases further and becomes negative 
on increasing $\lambda_k$ \cite{chattsen,goswami2}. The steps in those 
studies were taken as Right/Left according to whether it is Gain/Loss. 
In this study we are going to take a similar approach
for this CCM walk with a modification. Except for Gain and Loss, there is a third possibility.
When an agent 
 gains she/he moves a step towards Right and if she/he incurs a
loss moves a step towards Left. Apart from these two, the third possibility demands 
that the BTL tagged agent may not interact 
with another one if any one of them or both possesses a wealth less than $m_L$ and 
therefore the corresponding walker may stay put at its position in the GLS. 
The walks are correlated as when two
agents interact, if any one takes a right step, the other has to move towards left. Also if one is stay put, 
the other should stay put too.\\

\subsection{Measurement of bias}

For a lazy walker we know that it can have steps $1,0,-1$. It is obvious therefore that the CCM walk in this study can be
compared to a lazy walk. If the agent gains then the corresponding walker moves one step to 
the right, if loses, the walker moves towards left. If the interaction is missed due to either one of them 
or both falling below the threshold line, the walker remains in its 
position in the GLS. Just like earlier works, here also 
the amount of gain/loss is not important. \\

Consider a biased lazy walker with probability of going towards right $p_R$, towards left $p_L$ and probability to stay put 
$p_0$. Obviously $p_R+p_L+p_0=1$. The average distance traveled by such a walker is linear in $t$. Precisely, the average 
distance traversed can be written as
$\langle x \rangle = a(\lambda_k)t$. Here $a(\lambda_k)=[2p_R-(1-p_0)]$ is the slope of the line, 
a measure for the amount of drift. 
As in  any ballistic diffusion, here we have $(\langle x^2 \rangle - \langle x\rangle^2) \sim t^2$ for all $\lambda_k$ except  
when $\lambda_k \rightarrow \lambda_{k}^*$. For $\lambda_k \rightarrow \lambda_{k}^*$, we have observed that 
$(\langle x^2 \rangle - \langle x\rangle^2) \sim t$. 
In Fig. \ref{slopeAB}, the variation 
of $\langle x \rangle$ against $t$ is shown for $\lambda_k = 0.0,0.4$ and $0.9$ for the 
BTL tagged agent walker for both models A and B. Here we have taken $N = 256$ and $m_L=0.1$. 
Inset shows the plot of drift $a(\lambda_k)$ as a function of $\lambda_k$. The drifts are obtained from the slopes  
of $\langle x \rangle$ versus $t$ plot. 
At $\lambda_{k}^*$, for both the models we have found that $p_R=p_L$ and therefore drift $a(\lambda_{k}^*)=0$. 
The precise value of $\lambda_{k}^*$ is found to be 
$0.471$ for model A and $0.443$ for model B. 
For any $\lambda_k$ 
the slope is more positive for low $m_L$ and less positive for larger $m_L$. However, close to $\lambda_{k}^*$, the effect is 
almost negligible.
This is shown in Fig. 
\ref{slopevariation_mL}. This can be interpreted in the following way. The BTL agents' subsidy come 
from other agents' taxes, i.e., at the cost of others. As we increase $m_L$, number of BTL agents 
increase and the subsidy amount coming from the taxes of others increase. Therefore possibility of having an interaction 
decreases and the tendency to gain also decreases. As our definition for model B demands giving subsidy 
to the agent at every MC step, and that has to come from the others above the threshold line, therefore,  
the effect is more pronounced for model B compared to A. 
However, the amount of wealth possessed by a BTL agent increases as we 
increase $m_L$ for any specific $\lambda_k$. \\

\begin{figure}[!h]
\centering
\includegraphics[width=9cm,angle=-90]{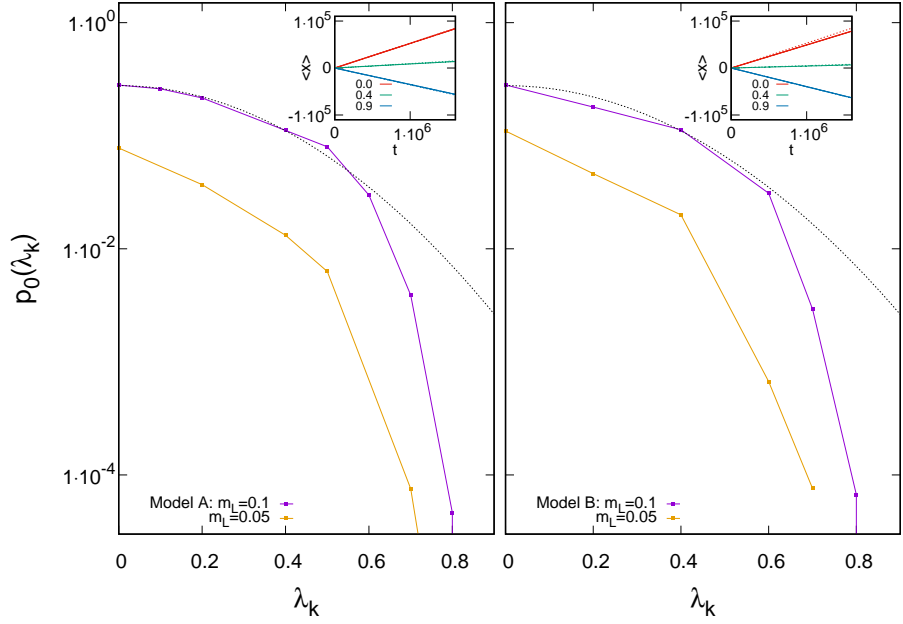}
\caption{Variation of $p_0(\lambda_k)$ against $\lambda_k$ for models A (Left) and B (Right) for $N = 256$ for $m_L=0.05$ (yellow), $0.1$ (violet). 
Black dotted line shows the nature of $a_0\exp(-b_0 x^2)$. 
Inset shows the comparison of slope obtained from $\langle x \rangle$ versus $t$ data (solid line) 
and using the parameters in lazy walk (dotted line).}
\label{slopeverific}
\end{figure}

We check the exact number of right, left and zero steps for the modified CCM walk and try to find out how 
the probabilities $p_R, p_L, p_0$ change with $\lambda_k$ for a specific $m_L$. As we know that 
$a(\lambda_k)=2p_R-(1-p_0)$, from Fig. \ref{slopeAB} it is clear that,
$p_R, p_L$ and $p_0$ are functions of $\lambda_k$. 
The specific probabilities for 
two $m_L$ values will be found 
in the Table \ref{p0pRpL}. For both the models the variation of $p_0(\lambda_k)$ against $\lambda_k$ 
are shown in Fig. \ref{slopeverific} 
for two different $m_L$. 
It is seen that the nature of variation matches well with the form 
$p_0(\lambda_k)\sim a_0\exp(-b_0 x^2)$ for both the models A and B, 
where $a_0$ and $b_0$ 
are two parameters, for low values of $\lambda_k$. The plots show discrepancy for high $\lambda_k$ values 
from the predicted behaviour. 
Now if we simulate a lazy 
walker with those $p_0, p_R$ and $p_L$ values that should show a similar $\langle x \rangle$ versus $t$ behaviour.  
In the inset of Fig.
\ref{slopeverific}, the $\langle x \rangle$ versus $t$ graph is compared with the same of a 
lazy walker for a few specific $\lambda_k$ and therefore $p_0, p_R$ and $p_L$ values values.\\

\begin{table}[!h]
\begin{center}
\caption{$p_0, p_R, p_L$ values for specified $m_L$ and $\lambda_k$ values for Model A and B 
for the modified CCM walker of BTL agent.}
\begin{tabular}{|c|c|c|c|c|c|c|c|}
\hline
$m_L$  &  $\lambda_k$   &   \multicolumn{3}{c|}{Model A}   & \multicolumn{3}{c|}{Model B} \\\cline{3-8}
   &  & $p_0$ &$p_R$ &$p_L$ & $p_0$ &$p_R$ &$p_L$ \\
\hline
0.1 & 0.0 & 0.285 & 0.384 & 0.331 & 0.282 & 0.386 & 0.332 \\
    & 0.4 & 0.111 & 0.449 & 0.439 & 0.113 & 0.446 & 0.441 \\
    & 0.9  & 0.0 & 0.482 & 0.518 & 0.0 & 0.480 & 0.520 \\
\hline
0.05 & 0.0 & 0.078 & 0.492 & 0.430 & 0.110 & 0.474 & 0.416 \\
     & 0.4 & 0.013 & 0.496 & 0.490 & 0.020 & 0.492 & 0.488 \\
     & 0.9 & 0.0   & 0.481 & 0.519 & 0.0 & 0.480 & 0.520 \\
\hline
\end{tabular}
\label{p0pRpL}
\end{center}

\end{table}

\subsection{Distribution of Path Lengths in the GLS}
We have seen in the previous section, how the probabilities $p_R, p_L, p_0$ changes with $\lambda_k$. To have a 
detailed understanding about the probabilities we are now going to study how the quantities vary with walk length $X$.
A path length $X$ here signifies the length traversed at a stretch without changing direction. For the Right/Left 
direction, this means the agent will gain/lose for $X$ steps continuously and after that it will either make a loss/gain 
or stay put. Here we study three such quantities $W_R(X), W_L(X)$ and $W_0(X)$ where the suffix indicates whether 
it is a gain or loss or no interaction. The distribution of path lengths in the GLS is an interesting quantity to study and 
was studied earlier in \cite{goswami2} where there were only Right/Left movements. 
Here, it is clear that:
\begin{equation}
W_i(X)\propto p_{i}^{X}(1-p_i)^2;  ~~~~~~i=0,R,L. \\
\end{equation}

\begin{figure}[!h]
\centering 
\includegraphics[trim=5cm 0cm 0cm 0cm,width=6cm, angle=-90,clip]{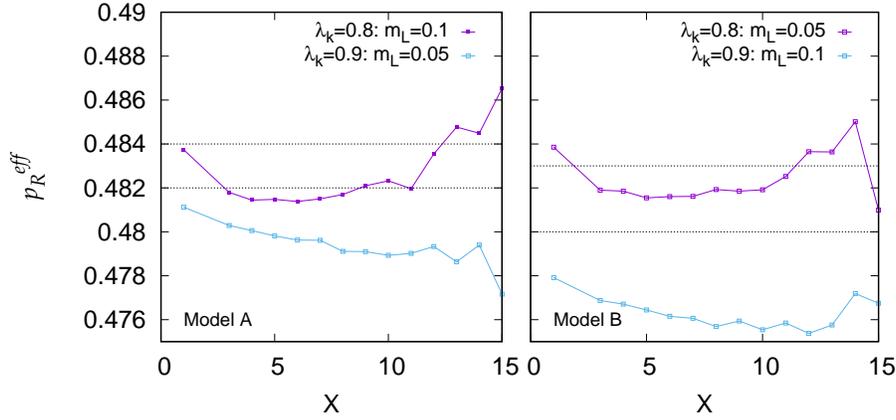}
\caption{Variation of $p_R^{eff}$ as a function of $X$ for models A ($\lambda_k=0.8, m_L=0.1$ (violet) and $\lambda_k=0.9, m_L=0.05$ (blue) shown) 
and B ($\lambda_k=0.8, m_L=0.05$ and $\lambda_k=0.9, m_L=0.1$ shown) for $N = 256$ obtained by distribution of walk lengths at a 
stretch. Black dotted lines show the corresponding $p_R$ values obtained from slopes.  
}
\label{walklength_pR}
\end{figure}

We now wish to extract the values of $p_R$ for a specific $\lambda_k$ and $m_L$ from the distribution of path lengths 
at a stretch. 
For high $\lambda_k$ values, e.g., $\lambda_k=0.8,0.9$, the probability 
$p_0$ is extremely small, and therefore the walk is very 
similar to a biased random walk. In that case, it is easy to extract some $p_R^{eff}(X, \lambda_k, m_L)$ as then 
we can approximately write
\begin{equation}
 \frac{W_R(X)}{W_L(X)}=(\frac{p_R}{1-p_R})^{X-2}.
 \label{compLR}
\end{equation}
$\frac{W_R(X)}{W_L(X)}$ is calculated numerically for a $\lambda_k$ and $m_L$.
The $p_R^{eff}$ is shown as a function of $X$ in Fig. \ref{walklength_pR} for a few $m_L$ values for both model A and B. 
The $p_R$ value has some variation over $X$ and not constant as predicted in Table \ref{p0pRpL}. 

For smaller $\lambda_k$ however we cannot use Eq. \ref{compLR}.
As for a lazy walker there are three parameters involved, we can use the obtained value of $p_0$ to check whether 
we are getting the same $p_R$ value as in Table \ref{p0pRpL} from this path length distribution data or not. 
For this we use the following: 
\begin{equation}
 \frac{W_R(X)}{W_0(X)}=\frac{p_{R}^{X}(1-p_R)^2}{p_{0}^{X}(1-p_0)^2}.
\end{equation}
For the BTL tagged agent's walk in GLS, we calculate $\frac{W_R(X)}{W_0(X)}$ numerically for specific values of 
$\lambda_k$ and $m_L$. 
The obtained $p_R^{eff}(X, \lambda_k, m_L)$ 
values do not match 
well except for the low values of $X$. 

\begin{figure}[!h]
\centering 
\includegraphics[width=9cm, angle=-90]{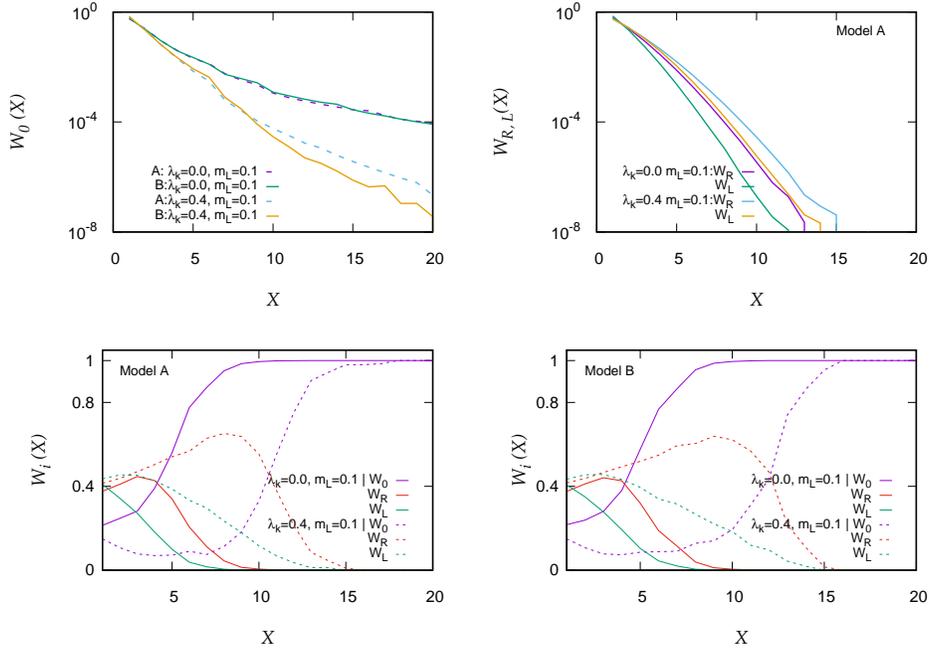}
\caption{Top Left : Variation of $W_0(X)$ against $X$ for models A and B for $\lambda_k=0.0, 0.4$ and $m_L=0.1$ (violet and blue for model A and green and yellow for model B). Top Right :  
Variation of $W_R(X)$ (violet and blue) and $W_{L}(X)$ (green and yellow) against $X$ for model A for $\lambda_k=0.0, 0.4$ and $m_L=0.1$. (Behaviour of the quantities 
for model B is similar as those of A and therefore it is not shown.)
Bottom Left : Relative variation of $W_0(X)$ (violet), $W_R(X)$ (red), $W_L(X)$ (green) against $X$ for model A for $\lambda_k=0.0, 0.4$ and $m_L=0.1$.
Bottom Right : Relative variation of $W_0(X)$ (violet), $W_R(X)$ (red), $W_L(X)$ (green) against $X$ for model B for 
$\lambda_k=0.0, 0.4$ and $m_L=0.1$.}
\label{WX_relall}
\end{figure}

From the above two aspects, therefore, we can say that the walker is not behaving like an usual biased lazy walker.
The variation of $W_0, W_R$ and $W_L$ as a function of $X$ are shown in Fig. \ref{WX_relall} a, b for A and B. 
As it can be seen the individual path distributions vary approximately as an exponential.
It is to be noted here that 
all the variations are shown such that $\sum_X W_i(X)=1$ where $i=0, R, L$. Also the relative variation of path distributions
$W_0, W_R$ and $W_L$ as a function of $X$ are shown in Fig. \ref{WX_relall} c, d, 
considering $W_0(X)+W_R(X)+W_L(X)=1$ for all 
$X$. It is clear from Fig. \ref{WX_relall} c, d that for large $X$,  
$W_R$ and $W_L$ decreases and finally comes to zero. That means long paths at a stretch 
for Gain/Loss are less probable. However, 
long $W_0$ paths are possible.

\begin{figure}[!h]
\centering
\includegraphics[trim=5cm 0cm 0cm 0cm,width=6cm, angle=-90,clip]{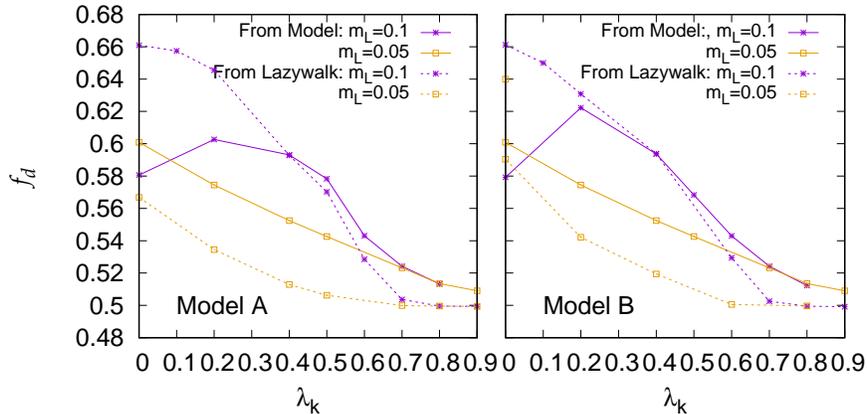}
\caption{Left : Direction Reversal Probability $f_d$ against $\lambda_k$ for model A for $m_L=0.1$ (violet) , $m_L=0.05$ (yellow). Solid lines indicate $f_d$ from distribution of free path and dashed lines indicate those from 
the simulated lazywalk. Right : Same for 
Model B}
\label{dirrev}
\end{figure}


Another interesting quantity to check here is the direction reversal probability. For lazy walker, the direction 
reversal probability is $2(p_0+p_R-p_0^2-p_R^2-p_0p_R)$. For our walk we consider the following quantity:
\begin{equation}
 \langle X \rangle=\sum_X (XW_R(X)+XW_L(X)+XW_0(X)).
\label{avgX}
\end{equation}
Here $\langle X \rangle$ is the average distance traveled in any particular direction (Gain, Loss or no movement) 
at a stretch. Obviously, the direction reversal probability is given by 
$f_{d}=\frac{1}{\langle X \rangle}$. 

The results are shown for some specific $\lambda_k$ and $m_L$ for A and B. It is seen that for our walker, probability 
is very high for all $\lambda_k$ values which saturates to a value close to $0.5$ for high $\lambda_k$. This is expected as 
when $\lambda_k$ is high, number of missed interactions become negligible. Therefore, the  
dynamics is controlled only through $p_R$ and the situation becomes similar as in case of 
Right/Left movement. These are shown in Fig. \ref{dirrev}. We also 
simulated a lazy walk using the parameters $p_0, p_R, p_L$ and calculated $f_d$ for that. 
For the lazy walker although the nature of variation is similar and tends to $0.5$ for high $\lambda_k$ 
as in case of our walker,
the exact probabilities do not match. 



\section{Correlation}
\begin{figure}[!h]
\centering
\includegraphics[height=.8\textwidth, angle=-90]{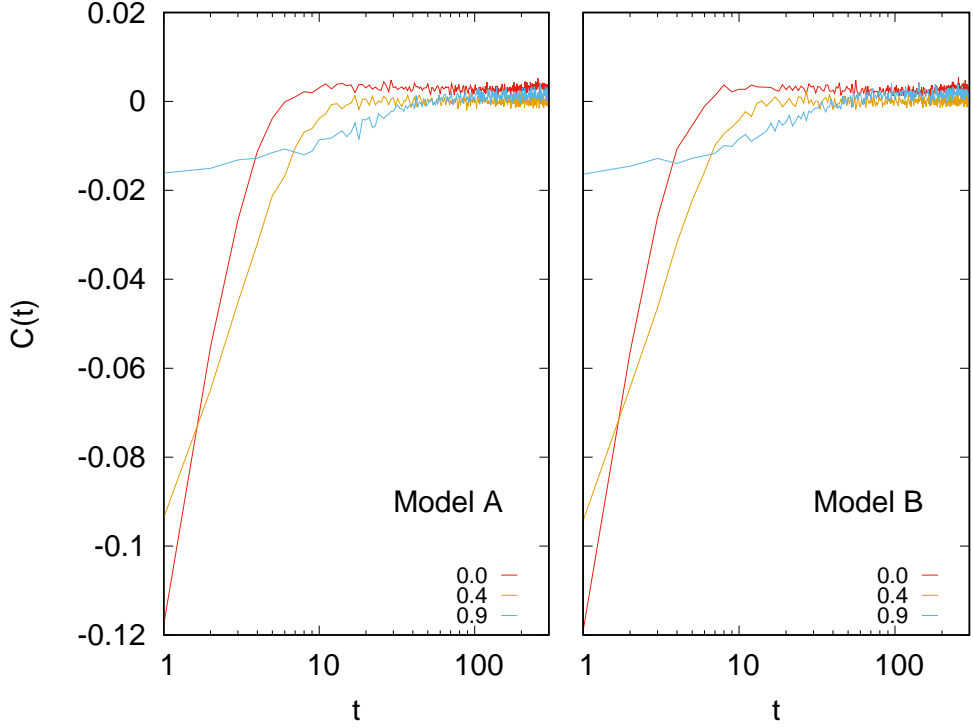}
\caption{Left : The correlation of steps $C(t)$ averaging over all possible initial
times $t_0$ for the walker for model A for $m_L=0.1$ and $\lambda_k=0.0$ (red), $0.4$ (yellow), $0.9$ (blue). Right : Same for Model B.}
\label{corr}
\end{figure}

We have seen that the steps of the walker have 
three possible values. Therefore, we need to analyze the time series for such a walk. 
Let the step taken at a time $t$ be written as  $s(t)=0, \pm 1$. The corresponding time correlation 
function can be written as:
$C(t)=\langle s(t_0) s(t_0+t)\rangle - s_0^2$
where $s_0^2=\langle s(t_0)\rangle \langle s(t_0+t)\rangle$. This can be written as we know that here 
$\langle s(t_0)\rangle$ is independent of $t$ and therefore $\langle s(t_0)\rangle=\langle s(t_0+t)\rangle=s_0$. 
Just as in CCM walk, here $\langle s_0 \rangle \neq 0$ unless near $\lambda_{k}^*$ for both the model A and B. The correlations for model 
A and B are shown in Fig. \ref{corr}. It is seen that there is a strong correlation near $t=1$ which gradually decreases 
with increasing $\lambda_k$.The short time correlations in both models A and B are negative. 
This  is consistent with the fact that direction
reversal probability is greater than $1/2$.  However, for one particular $\lambda_k$ the correlation ultimately 
saturates to a value $C_{sat}$ when $t\rightarrow \infty$. 

\begin{figure}[!h]
\includegraphics[trim=5cm 0cm 0cm 0cm,width=6cm, angle=-90,clip]{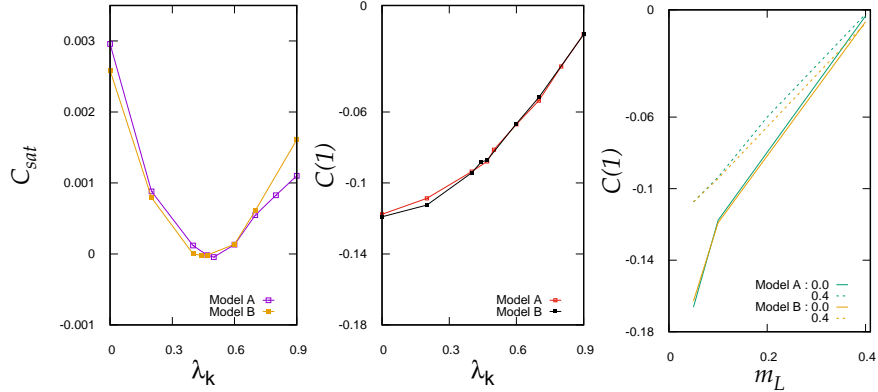}
\caption{Left : The saturation value of correlation of steps $C_{sat}$ against $\lambda_k$ for $m_L=0.1$ for A (violet) and B (yellow). 
Middle : The strongest correlation $C(1)$ as a function of $\lambda_k$ for $m_L=0.1$ for the models A (red) and B (black). 
Right : $C(1)$ against $m_L$ for A (green) and B (yellow) for $\lambda_k=0.0$ (solid line) and $\lambda_k=0.4$ (dashed line).}
\label{c1csat}
\end{figure}

This kind of feature was earlier noticed in \cite{goswami2}. 
The saturation value $C_{sat}$ is estimated by averaging near the end over a few hundred values of $t$. $C_{sat}$ as a function 
of $\lambda_k$ 
is shown for $m_L=0.1$ for both models A and B in Fig. \ref{c1csat} a. It is seen that 
 $C_{sat}$ reaches a minimum close to $\lambda_{k}^*$ which is $0.471$ for model A and $0.443$ for model B. 
 The minimum value of 
 $C_{sat}$ as observed for a $\lambda_k \approx \lambda_{k}^*$ is $\sim O(10^{-5})$. Also as we observe 
the strongest correlation $C(1)$ is changing with $\lambda_k$, we plot the same in Fig. \ref{c1csat} b. 
It is observed that as we 
increase threshold line $m_L$, the correlation over one step $C(1)$ becomes weaker and weaker for lower $\lambda_k$ values. 
However for high $\lambda_k$ for all $m_L$ values, $C(1)$ is almost a constant. 
This feature is similar in model A and B  but for higher values of $m_L$ the strength of maximum correlation, i.e., $C(1)$
is weaker for a particular $m_L$ in model $B$. This is shown in Fig. \ref{c1csat} c. 

\section{Reason Behind High Direction Reversal}
It is already seen that the probability of direction reversal in the GLS is very high. This signature is also clear from 
the strong correlation for small $t$. In this part, we are going to analyze why this direction reversal is preferred 
by an agent in this walk. \\


For our 
convenience, we choose the DY model which is the simplest among all and the form of wealth distribution is well known. 
For this we mimic a situation when the agent ends up with gain and then again interacts. The probability that 
the agent will incur a loss in the next step requires that she/he has to interact with another with low value of wealth. We 
consider our agent ends up with wealth $m'$ in the first step. 
However, there are two possibilities in this case. If the second agent's wealth is between $0$ and $m_L$ then 
there will be no interaction at all and if that is higher than $m_L$ and less than $m'$, our agent may lose some. 
The conditional probability that the agent will have a loss after a gain is
\begin{equation}
 W'(LG)=\frac{\int_{m_1}^{\infty} P(m_1)dm_1 \int_{m_L}^{m_1} P(m_2) dm_2}{\int_{m_1}^{\infty} P(m_1)dm_1}
\end{equation}
and the probability that the interaction will not occur is
\begin{equation}
 W'(0G)=\frac{\int_{m_1}^{\infty} P(m_1)dm_1 \int_{0}^{m_L} P(m_2) dm_2}{\int_{m_1}^{\infty} P(m_1)dm_1}
\end{equation}
But the probability that the previous step was a gain is $p_R$. Considering the wealth distribution of the form 
$P(m)\sim \exp(-m)$, the probability that it will either lose or stay put is 
$W(iG)=p_R(\lambda_k)\times(1-\frac{1}{2}\exp(-m_1))$ where $i=L/0$.\\

Similarly the conditional probability of having a Loss in the first step and then either a Gain or a no interaction is 
\begin{equation}
 W'(jL)=\frac{1}{2}[1+\exp(-m_1)]+1-\exp(-m_L)
\end{equation}
where $j=G/0$.
Therefore $W(jL)=p_L(\lambda_k)W'(jL)=p_L(\lambda_k)[\frac{1}{2}[1+\exp(-m_1)]+1-\exp(-m_L)]$.\\

Proceeding in a similar manner we can show that 
\begin{equation}
 W(i0)=p_0(\lambda_k)
\end{equation}
where $i=G/L$. \\

Thus, the probability of direction reversal is 
\begin{equation}
 f_d(\lambda_k)=p_0(\lambda_k)+p_R(\lambda_k)(1-\frac{1}{2}\exp(-m_1))+p_L(\lambda_k)[\frac{1}{2}[1+\exp(-m_1)]+1-\exp(-m_L)].
\end{equation}
In any case we can show that 
\begin{equation}
\label{fd}
 f_d(\lambda_k) > p_0(\lambda_k)+\frac{1}{2}p_R(\lambda_k)+p_L(\lambda_k)[\frac{3}{2}-\exp(-m_L)]
\end{equation}
Using the obtained values of $p_0, p_R$ and $p_L$ for different $\lambda_k$ and $m_L$ we check 
the lower bound of the direction reversal probabilities from the  Eq. \ref{fd}. $f_d$ for a few $\lambda_k$ and 
$m_L$ values obtained using Eq. \ref{fd} and from the model are compared in Table \ref{table2}.

\begin{table}[!h]
\begin{center}
\caption{Comparison of direction reversal probabilities $f_d$ obtained from Eq. \ref{fd} and from the model A}
\begin{tabular}{|c|c|c|c|c|}
\hline
$m_L$  &  $\lambda_k$ &   \multicolumn{3}{c|}{$f_d$} \\\cline{3-5}
   &  & From Eq. \ref{fd} & From Lazywalk & From model \\
\hline
  0.1 & 0.0 & 0.672  & 0.661 & 0.580\\
    & 0.4 & 0.598 & 0.593 & 0.593 \\
    & 0.9  & 0.548 & 0.499 & 0.508 \\
\hline
0.05 & 0.0 & 0.560 & 0.567 & 0.600 \\
     & 0.4 & 0.530 & 0.513 & 0.552 \\
     & 0.9 & 0.525 & 0.499 & 0.508\\
\hline
\end{tabular}
\label{table2}
\end{center}
\end{table}

As we can see from Table \ref{table2}, there is good agreement of the lower bound of the inequality with the lazy walker 
except when $\lambda_k$ is very high. Also the agreement is not that good for the data obtained from our model. This 
indicates that the walk of our agent is not exactly a lazy walk.
As $\lambda_k$ becomes very high the DY model approximation is no longer valid and therefore there is discrepancy from 
the direction reversal probability obtained from Eq. \ref{fd}.

\section{Summary and Conclusion}
In this work, the nature of transactions
made in CCM model with some modification is studied. We used the idea of threshold line below which an agent is 
identified as a BTL one at the time of assigning wealth. 
These agents are always eligible for subsidy. This is similar to giving opportunity to 
some backward class people in any caste-based system.
The threshold line is important in another way. 
It dictates whether an interaction will occur or not. Any agent, either BTL or not, having 
wealth below that line at any point of interaction is considered insecure as in 
any real situation. If either one or both the interacting agents' wealth is below the 
threshold line, the interaction will not occur. We also considered
an equivalent picture of a $1$D walk in an abstract space for Gains and Losses. Here amount of Gain/Loss is not important; 
we have just used the information whether it is a Gain or Loss. As a tagged agent may have Gain, Loss or no interaction, 
the corresponding walker is a lazy walker which in addition to 
Left/Right movement, may stay put at a position. The high direction reversal probability indicates 
that there is a high tendency of individuals to make
a gain or to stay put immediately after a loss and vice versa. This kind of effect was studied for usual CCM model before. 
Here also we found this to be compatible with
human psychology. From the high direction reversal probability it is clear that after a 'no' interaction the agent will 
try to interact immediately at the next step. Also a person may take part in an interaction which may lead either to a loss 
or to no interaction, when she/he had a
gain in the previous step. After suffering a loss, in a same manner, a person 
will either try to have a gain or may stay put due to 
insecurity. This
effect is maximum for zero saving propensity and decreases with increased saving. The data obtained 
from correlation for one step 
also indicates that there will be high tendency of gain after loss and vice versa. (of course, there may be stay 
puts in between). \\

The subsidy is given here in two ways, firstly at the time of assigning the wealth to the agents initially (Model A), 
and secondly, 
at each MC step (Model B). 
It is seen that if subsidy is given repeatedly (i.e., at each MC step), 
the agent moves with a less positive drift in the GLS for Model B compared 
to Model A for any particular 
$\lambda_k$ (Fig. \ref{slopeAB}). This can be understood, as, giving repeated subsidy to a BTL agent will affect the wealth of others 
and as the walks are actually correlated, it affects the tagged BTL walk in turn. The amount of wealth possessed by such an agent 
is greater in Model B compared to Model A. The parameters $p_0, p_R$ and $p_L$ are 
obtained from the walk of the tagged BTL agent and with those a lazy walk is simulated. It is seen that the BTL agent's  
CCM-like walk is not exactly similar to a biased lazy walk. The walk has no bias for some saving propensity $\lambda_{k}^*$ which is 
different for models A and B. \\

The distribution of path traveled at a stretch $X$ is studied for Right/Left movements and staying put. 
The quantities are $W_R(X), W_L(X)$ and $W_0(X)$ where the suffix indicates whether 
it is a gain or loss or no interaction.
Using that we calculated the direction reversal probability $f_d$ and the same is calculated analytically for a DY model. 
Using the lazy walker 
parameters we compared the analytical direction reversal probability with those obtained from the walk. The analytical 
approach indicates that $f_d(\lambda_k)$ must be greater than 
$p_0(\lambda_k)+\frac{1}{2}p_R(\lambda_k)+p_L(\lambda_k)[\frac{3}{2}-\exp(-m_L)]$. 
However, the analytical treatment was done using the 
DY model wealth distribution and therefore is not in good agreement to our case, when saving propensity is high. 
From different calculations shown here it is clear that for Model A, $\lambda_k=\lambda_{k}^*\simeq 0.471$ 
and for model B it is close to $0.443$. Although for Model A it is close to 
the $\lambda_{k}^*$ obtained for usual CCM model, i.e., $0.469$, for Model B the result is different.

Finally, our study shows a modified CCM model with the concept of a Threshold line.   
The threshold line in addition to identification of BTL agents, puts restriction to interactions. 
We have seen that putting restriction changes the distribution which is dependent on the threshold line $m_L$ as 
shown in Fig. \ref{mnydist_compare}.
The dynamics of a tagged BTL agent is compared to a Lazy walker in GLS. 
The high direction reversal probability for such a walker indicates that
one can afford to have a loss or may stay put he/she has gained begore. Also after suffering a loss, it may 
stay put or may try to make a gain.
This is completely compatible with 
human psychology. 
Also, it is seen that the value of $\lambda_{k}^{*}$ is 
independent of $m_L$ for both the models A and B as shown in Fig. \ref{slopevariation_mL}. 
The effect of giving subsidy once over giving repeated subsidy 
to such agents 
is checked from the point of view of the walk in GLS. The 
walk is seen to have more positive drift when the subsidy is given once.








\end{document}